\documentclass[hyper,11pt,letterpaper]{JHEP3}
\usepackage[dvips]{epsfig}
\usepackage{amsmath,amssymb,epsf,amsfonts}
\usepackage{cite}
\usepackage{graphicx}
\usepackage{dcolumn}
\usepackage{bm}

\def\Ti{\langle T^t_{~i} \rangle}
\def\Ui{U^t_{~i}}
\def\N{{\mathcal{N}}}

\def\mink{\mathbb{R}^{3,1}}
\def\<{\langle}

\def\>{\rangle}
\def\({\left (}
\def\){\right )}
\def\[{\left[}
\def\]{\right]}

\def\beq{\begin{equation}}
\def\eeq{\end{equation}}
\def\gx{g_{xx}}
\def\gt{|g_{tt}|}
\def\gu{g_{uu}}
\def\E{\tilde{E}}
\def\Bz{\tilde{B}_z}
\def\Bx{\tilde{B}_x}
\def\EB{\vec{E} \cdot \vec{B}}
\def\F{\tilde{F}}
\def\at{\tilde{A}'_t}
\def\atsq{\tilde{A}_t^{'2}}
\def\ax{\tilde{A}'_x}
\def\axsq{\tilde{A}_x^{'2}}
\def\ay{\tilde{A}'_y}
\def\aysq{\tilde{A}_y^{'2}}
\def\az{\tilde{A}'_z}
\def\azsq{\tilde{A}_z^{'2}}

\def\jt{\langle J^t \rangle}
\def\jx{\langle J^x \rangle}
\def\jy{\langle J^y \rangle}
\def\jz{\langle J^z \rangle}

\def\p{\pi}                
\def\a{\alpha}
\def\s{\sigma}
\def\lam{\lambda}

\newcommand{\bea}{\begin{eqnarray}}
\newcommand{\eea}{\end{eqnarray}}
\def\sxx{\sigma_{xx}}
\def\sxy{\sigma_{xy}}
\def\sxz{\sigma_{xz}}

\def\pa{(2 \pi \alpha')}

\def\Om{{\cal{O}}_m}
\def\Omv{\langle {\cal{O}}_m \rangle}
\def\Nf{N_f}
\def\Nc{N_c}
\def\rarrow{\rightarrow}
\def\ra{\rightarrow}

\def\a{\alpha}

\def\g{\gamma}

\def\lam{\lambda}
\def\m{\mu}
\def\n{\nu}
  
\def\p{\pi}                
\def\s{\sigma}                                   

\title{\LARGE Holographic Flavor Transport in Arbitrary Constant Background Fields}
\author{
Martin Ammon\footnotemark[1] , Thanh Hai Ngo\footnotemark[2] , and Andy O'Bannon\footnotemark[3]
\\Max Planck Institut f\"{u}r Physik (Werner Heisenberg Institut) 
\\ F\"{o}hringer Ring 6, 80805 M\"{u}nchen, Germany}

\footnotetext[1]{E-mail: \email{ammon@mppmu.mpg.de}}
\footnotetext[2]{E-mail: \email{ngo@mppmu.mpg.de}}
\footnotetext[3]{E-mail: \email{ahob@mppmu.mpg.de}}

\abstract
{
We use gauge-gravity duality to compute a new transport coefficient associated with a number $\Nf$ of massive $\N=2$ supersymmetric hypermultiplet fields propagating through an $\N=4$ $SU(N_c)$ super-Yang-Mills theory plasma in the limits of large $N_c$ and large 't Hooft coupling, with $\Nf \ll \Nc$. We introduce a baryon number density as well as arbitrary constant electric and magnetic fields, generalizing previous calculations by including a magnetic field with a component parallel to the electric field. We can thus compute all components of the conductivity tensor associated with transport of baryon number charge, including a component never before calculated in gauge-gravity duality. We also compute the contribution that the flavor degrees of freedom make to the stress-energy tensor, which exhibits divergences associated with the rates of energy and momentum loss of the flavor degrees of freedom. We discuss two currents that are free from these divergences, one of which becomes anomalous when the magnetic field has a component parallel to the electric field and hence may be related to recent study of charge transport in the presence of anomalies.
}
\preprint{MPP-2009-154}

\begin{document}

\section{Introduction}

The conductivity tensor $\s_{ij}$ measures the electrical response of a conducting medium to externally applied fields. It is defined by
\beq
\< J_i \> = \s_{ij} \, E_j, \nonumber
\eeq
where $E$ are externally applied electric fields and $\< J_i \>$ are the electrical currents induced in the medium. Similarly, the thermoelectric conductivity tensor $\alpha_{ij}$ measures the thermal response. It is defined as
\beq
\< Q_i \> = \alpha_{ij} \, E_j, \nonumber
\eeq
where $\< Q_i \>$ are heat currents induced in the medium,
\beq
\< Q_i \> = \< T^{t}_{~i} \> - \mu \< J_i \>, \nonumber
\eeq
where $\< T^i_{~j} \>$ are the components of the stress-energy tensor, so that $\< T^t_{~i} \>$ are momentum densities, and $\mu$ is the chemical potential.

Our goal in this paper is to use the anti-de Sitter / Conformal Field Theory correspondence (AdS/CFT) \cite{Maldacena:1997re,Gubser:1998bc,Witten:1998qj} to compute a conductivity tensor and the contribution to the stress-energy tensor associated with a number $N_f$ of massive $\N = 2$ supersymmetric hypermultiplet fields propagating through an ${\cal N }=4$ supersymmetric $SU(N_c)$ Yang-Mills (SYM) theory plasma at temperature $T$. We will study hypermultiplet fields that transform in the fundamental representation of the gauge group, \textit{i.e.} flavor fields. We will work in the limits $\Nc \rarrow \infty$ with the 't Hooft coupling $\lambda \equiv g_{YM}^2 \Nc$ fixed, and with $\lambda \gg 1$. We also fix $\Nf$ such that $\Nf \ll \Nc$, and work to leading order in $\Nf/\Nc$.

AdS/CFT equates the SYM theory in the limits above with supergravity on the ten-dimensional spacetime $AdS_5 \times S^5$, where $AdS_5$ is (4+1)-dimensional anti-de Sitter space and $S^5$ is a five-sphere \cite{Maldacena:1997re}. We should imagine that the SYM theory ``lives'' on the boundary of the $AdS_5$ space, so in this sense AdS/CFT is ``holographic.'' The SYM theory in thermal equilibrium is dual to supergravity on an AdS-Schwarzschild spacetime, where the SYM theory temperature $T$ is identified with the Hawking temperature of the AdS-Schwarzschild black hole \cite{Witten:1998zw,Gubser:1996de}. The $\Nf$ hypermultiplets appear in the supergravity description as a number $\Nf$ of D7-branes embedded in the AdS-Schwarzschild background \cite{Karch:2002sh}. With $\Nf \ll \Nc$ D7-branes, we may treat the branes as probes and neglect their back-reaction on the supergravity fields.  We will explain the supergravity description of the hypermultiplet mass and the electric and magnetic fields in the sequel. Though we focus on this system, our analysis easily extends to other probe D-brane systems.

Our work is part of a larger program of studying transport phenomena in strongly-coupled systems using gauge-gravity duality, which provides many solvable toy models of such systems. The study of these toy models has provided qualitative (and often quantitative) insight into real physical systems, including the quark-gluon plasma created at the Relativistic Heavy-Ion Collider as well as various systems in condensed matter physics, especially systems whose low-energy dynamics is controlled by a nearby quantum critical point.

When the $N_f$ flavors have equal masses, the theory has a global $U(N_f)$ symmetry, whose $U(1)_B$ subgroup we identify as baryon number (hence the subscript). We will denote the $U(1)_B$ current as $J^{\mu}$. We will study the theory with a finite $U(1)_B$ density $\jt$. We will also introduce static external electric and magnetic fields that couple to anything with $U(1)_B$ charge. For \textit{perpendicular} electric and magnetic fields, the conductivity for this system was computed in refs. \cite{Karch:2007pd,O'Bannon:2007in}, while the contribution that the flavor fields make to the stress-energy tensor was computed in ref. \cite{Karch:2008uy}.

We will generalize the results of refs. \cite{Karch:2007pd,O'Bannon:2007in,Karch:2008uy} to completely arbitrary (constant) electric and magnetic fields. For an arbitrary configuration of constant electric and magnetic fields, we may sum all the electric fields into a single vector, and similarly for the magnetic fields. The most general configuration is thus an electric field $\vec{E}$ pointing in some direction, which we will take to be $\hat{x}$, and a magnetic field $\vec{B}$ that may be decomposed into two components, one along $\hat{x}$, which we call $B_x$, and one perpendicular to it, along the $\hat{z}$ direction, which we call $B_z$. Stated simply, then, we will generalize the results of refs. \cite{Karch:2007pd,O'Bannon:2007in,Karch:2008uy} to include a magnetic field with nonzero $\hat{x}$ component, or equivalently a nonzero $\EB \sim F \wedge F$.

Introducing a nonzero $\EB$ is worthwhile for a number of reasons\footnote{The authors of ref. \cite{Lifschytz:2009si} studied transport of baryon number charge in the presence of nonzero $\EB$ in a holographic model of Quantum Chromodynamics, the Sakai-Sugimoto model \cite{Sakai:2004cn}. They found that the system behaves as a perfect conductor. The crucial ingredient there was the anomaly in the axial $U(1)$ current, however, which is absent in our system (more precisely, the effects of the anomaly are suppressed in the probe limit).}. With perpendicular electric and magnetic fields $E$ and $B_z$, we expect a current $\jx$ parallel to the electric field (because it pushes the charges) and a Hall current $\jy$ orthogonal to both the electric and magnetic field. With nonzero $B_x$, we expect a current $\jz$, and hence we can compute a new transport coefficient, $\s_{xz}$. More generally, we can compute the entire conductivity tensor and determine its dependence on $B_x$. As mentioned in refs. \cite{Karch:2007pd,O'Bannon:2007in,Karch:2008uy}, two types of charge carriers contribute to the currents. The system has not only the charge carriers we introduce explicitly in $\jt$, but also charge carriers coming from pair-production in the external electric field. We find that, generically, $B_x$ enhances the contribution from the pair-produced charges.

Additionally, in a Lorentz-invariant system, we can build two Lorentz-invariant quantities from $\vec{E}$ and $\vec{B}$, namely $|\vec{E}|^2 - |\vec{B}|^2$ and $\EB$. When $\EB = 0$, and $|\vec{B}| > |\vec{E}|$, we can boost to a frame where the electric field is zero, which immediately tells us that all the physics must be equilibrium. For example, as reviewed in ref. \cite{Hartnoll:2007ai}, the form of the Hall conductivity is fixed by Lorentz invariance to be $\jt/B_z$. When $\EB$ is nonzero we can no longer boost to a frame in which the electric field is zero, hence the physics cannot be purely equilibrium.

Many (though not all) previous gauge-gravity calculations of conductivities were in (3+1)-dimensional AdS space, so that the boundary CFT was (2+1)-dimensional \cite{Herzog:2007ij,Hartnoll:2007ai,Hartnoll:2007ih,Hartnoll:2007ip}, which precludes the existence of $\EB$. Another drawback of the systems studied in refs. \cite{Herzog:2007ij,Hartnoll:2007ai,Hartnoll:2007ih,Hartnoll:2007ip} was translation invariance, which implies momentum conservation. The system thus has no way to dissipate momentum, so the DC transport behavior was singular. For example, the DC conductivity at finite density is infinite because the charge carriers, in the presence of an external electric field but without frictional forces, accelerate forever.

The probe limit $N_f \ll N_c$ effectively provides a mechanism for dissipation. (A more accurate description may be that the probe limit allows our system to mimic a dissipative system.) As explained in more detail in refs. \cite{Karch:2007pd,O'Bannon:2007in,Karch:2008uy,O'Bannon:2008aj}, and as we will review below, the charge carriers do indeed transfer energy and momentum to the $\N=4$ plasma, but the \textit{rates} at which they do so are of order $N_c$. That means that only at times of order $N_c$ will the charge carriers have transferred order $N_c^2$ amounts of energy and momentum to the plasma, and hence the motion of the $\N=4$ SYM plasma will no longer be negligible. For earlier times, we may treat the $\N=4$ SYM plasma as a motionless reservoir into which the charge carriers may dump their energy and momentum, thus providing the charge carriers with an (apparent) mechanism for dissipation.

As first demonstrated in ref. \cite{Karch:2008uy}, we can compute holographically the rates at which the charge carriers lose energy and momentum. To do so, we compute the contribution that the charge carriers make to the stress-energy tensor of the SYM theory. The loss rates appear in two places. First, the holographic results for the energy and momentum densities exhibit divergences whose coefficients (using a suitable regulator) we can identify as the loss rates. Second, the loss rates appear explicitly as components of the holographic result for the stress-energy tensor, namely components whose upper index is the holographic (radial) coordinate \cite{Karch:2008uy,Gubser:2008vz}.

Furthermore, as in ref. \cite{Karch:2008uy}, we will study observers in the field theory who ``see'' no loss rates. The simplest example is an observer who moves along with the charges: in that frame, the charges are at rest, so obviously such an observer should not see the charges lose energy and momentum. As in ref. \cite{Karch:2008uy}, we can confirm that our holographic result correctly reproduces the absence of loss rates. As mentioned in ref. \cite{Karch:2008uy}, we can also find a second observer who sees no loss rates, but only when $\EB=0$. When $\EB \neq 0$, the observer measures a current with nonzero divergence $ J^2 \, (\EB)$, where $J^2 = \langle J_{\mu} \rangle \langle J^{\mu} \rangle$. The identity of this observer was left as an open question in ref. \cite{Karch:2008uy}. Here we find that this observer's four-vector is in fact the magnetic field as measured by the moving charges. Much like the holographic result for the stress-energy tensor, the loss rate $J^2 \, (\EB)$ appears as the coefficient of a (suitably regulated) divergence in the current itself. Notice also that, given the $\EB$ anomaly in this current, if we were to study transport of the charge associated with this current we should find a special kinetic coefficient \cite{Erdmenger:2008rm,Banerjee:2008th} whose form is fixed by the anomaly coefficient (in our case, $J^2$) and thermodynamics (the equation of state), as explained in ref. \cite{Son:2009tf}.

This paper is organized as follows. In section \ref{d7solution} we present a solution for the worldvolume fields of probe D7-branes in the AdS-Schwarzschild background, representing a finite baryon density of flavor degrees of freedom in the presence of external electric and magnetic fields. In section \ref{conductivity} we use our gravity solution to compute the conductivity tensor associated with transport of baryon number charge. In section \ref{stressenergy} we compute the contribution that the flavor degrees of freedom make to the stress-energy tensor, study divergences in the components of the stress-energy tensor and their relation to energy and momentum loss rates, and then discuss two reference frames in which the divergences are absent. We conclude with some suggestions for future research in section \ref{conclusion}. We collect some technical results in an appendix.

\section{The Probe D7-brane Solution}
\label{d7solution}

In this section we present a solution of supergravity, plus probe D7-branes, describing massive hypermultiplets propagating through an $\N=4$ SYM plasma with finite $U(1)_B$ density and in the presence of external electric and magnetic fields.

The supergravity solution includes a ten-dimensional metric with a (4+1)-dimensional AdS-Schwarzschild factor and an $S^5$ factor. We will use an AdS-Schwarzschild metric
\beq
ds^2_{AdS_5}  = g_{uu} \, du^2 + \, g_{tt} \, dt^2 + g_{xx} \, d\vec{x}^2,
\eeq
where $u$ is the AdS radial coordinate. When we need an explicit metric, we will use
\beq
ds_{AdS_5}^2 = \frac{du^2}{u^2} - \frac{1}{u^2} \frac{(1 - u^4 / u_H^4)^2}{1+u^4/ u_H^4} \, dt^2 + \frac{1}{u^2} (1+u^4 / u_H^4) \, d\vec{x}^2.
\eeq
The boundary is at $u=0$ and the horizon is at $u = u_H$ with $u_H^{-1} = \frac{\p}{\sqrt{2}} \, T$. Here we are using units in which the radius of AdS is equal to one. In these units, we convert from string theory to SYM theory quantities using $\alpha'^{-2} = \lambda$. We will use an $S^5$ metric of the form
\beq
ds^2_{S^5} = d\theta^2 + \sin^2 \theta \, ds^2_{S^1} + \cos^2 \theta \, ds^2_{S^3},
\eeq
where $\theta$ is an angle between zero and $\pi/2$ and $ds^2_{S^1}$ and $ds^2_{S^3}$ are metrics for a unit-radius circle and 3-sphere, respectively. The supergravity solution also includes $N_c$ units of five-form flux through the $S^5$, but the five-form will be irrelevant in what follows, so we omit it.

We next introduce $N_f$ coincident probe D7-branes. As we will be interested only in the $U(1)$ part of the $U(N_f)$ worldvolume gauge field, the relevant part of their action will be the Dirac-Born-Infeld (DBI) term,
\beq
\label{originaldbi}
S_{D7} = - N_f T_{D7} \int d^8 \zeta \sqrt{-det\left(g_{ab} + (2\pi\alpha') F_{ab} \right)}.
\eeq
Here $T_{D7}$ is the D7-brane tension, $\zeta^a$ are the worldvolume coordinates, $g_{ab}$ is the induced worldvolume metric, and $F_{ab}$ is the $U(1)$ worldvolume field strength. The D7-branes will be extended along all of the $AdS_5$ directions, as well as the $S^3$ directions inside the $S^5$.

Our ansatz for the worldvolume fields will include the worldvolume scalar $\theta(u)$. The D7-brane induced metric is then identical to the background metric, except for the radial component, which is $g_{uu} = \frac{1}{u^2} + \theta'(u)^2$, where prime denotes differentiation with respect to $u$. Starting now, the notation $g_{uu}$ will include the $\theta'^2$ term. We will discuss $\theta(u)$'s equation of motion, boundary conditions, and holographic dual operator later in this section.

The $U(N_f)$ gauge invariance of the coincident D7-branes is dual to the $U(N_f)$ symmetry of the mass-degenerate flavor fields in the SYM theory. We identify the $U(1)$ subgroup as baryon number, $U(1)_B$. The D7-brane worldvolume Abelian gauge field $A_{\mu}$ is dual to the SYM $U(1)_B$ current $J^{\mu}$, so to introduce a finite $U(1)_B$ density in the SYM theory, we must introduce the worldvolume gauge field $A_t(u)$.

To introduce electric and magnetic fields, and the resulting currents $\jx$, $\jy$, and $\jz$, we also include in our ansatz the gauge field components
\beq
A_x(t,u) = -Et + f_x(u), \qquad A_y(x,u) = B_z \, x + f_y(u), \qquad A_z(y,u) = B_x \, y + f_z(u).
\eeq
In each case, the leading term is a non-normalizable mode that introduces an external field into the SYM theory. Choosing a gauge in which $A_u = 0$, we can write the nonzero elements of $F_{ab}$:
\beq
F_{tx} = - E, \qquad F_{xy} = B_z, \qquad F_{yz} = B_x,
\eeq
\beq
F_{ut} = A'_t, \qquad F_{ux} = A'_x, \qquad F_{uy} = A'_y, \qquad F_{uz} = A'_z.
\eeq

We will now write the action for our ansatz. Let us first define some notation. The fields in our ansatz depend only on $u$, so in eq. (\ref{originaldbi}) we can immediately perform the integration over the SYM theory directions $(t,x,y,z)$ and over the $S^3$ directions. Starting now we will divide both sides of eq. (\ref{originaldbi}) by the volume of $\mink$, so $S_{D7}$ will actually denote an action density. $L$ will denote the Lagrangian density, $S_{D7} \equiv - \int du \, L$. Using $T_{D7} = \frac{\alpha'^{-4} g_s^{-1}}{(2\pi)^7} = \frac{\lambda N_c}{2^5 \pi^6}$, we will also define the constant
\beq
\N \equiv N_f T_{D7} V_{S^3} = \frac{\lambda}{(2\pi)^4} N_f N_c,
\eeq
where $V_{S^3} = 2 \pi^2$ is the volume of a unit-radius $S^3$. Lastly, a tilde over a quantity denotes a factor of $\pa$, for example, $\tilde{F}_{ab} \equiv \pa F_{ab}$.

Plugging our ansatz into the action eq. (\ref{originaldbi}), we have
\beq
\label{actionD7}
S_{D7} = - \N \, \int du \, \cos^3 \theta \, \sqrt{\gu \gt \gx^3 - \gx A_2 - A_4},
\eeq
where $A_2$ and $A_4$ contain terms with two or four factors of $\tilde{F}_{ab}$, respectively,
\beq
\label{actiona2def}
A_2  = \gu \gx \E^2 + g_{tt} \gu (\Bx^2 + \Bz^2) + \gx^2 \atsq + g_{tt}g_{xx} \left( \axsq + \aysq + \azsq \right),
\eeq
\bea
\label{actiona4def}
A_4 & = & \gx \E^2 \left( \aysq + \azsq \right) + \gx \atsq \left( \Bx^2 + \Bz^2 \right) + \gu \E^2 \Bx^2 + g_{tt} \Bz^2 \azsq + g_{tt} \Bx^2 \axsq \\ & & +  2 g_{tt} \Bx \Bz \ax \az - 2 \gx \E \Bz \at \ay. \nonumber
\eea

The action only depends on the $u$ derivatives of $A_t$, $A_x$, $A_y$, and $A_z$, so the system has four constants of motion. As shown in refs. \cite{Karch:2007pd,O'Bannon:2007in}, we can identify these as the components of the $U(1)_B$ current density in the SYM theory\footnote{As in refs. \cite{Karch:2007pd,O'Bannon:2007in}, the D7-brane action diverges due to integration all the way to the $AdS_5$ boundary at $u=0$,  and thus requires renormalization. The recipe for the ``holographic renormalization'' of the D7-brane action appears in refs. \cite{Karch:2005ms,Karch:2006bv,Karch:2007pd,O'Bannon:2007in}. We first introduce a cutoff at $u=\epsilon$ and then add a counterterm action $S_{CT}$ to cancel the divergences as $\epsilon \ra 0$. The precise expression for $\langle J^{\mu} \rangle$ is
\beq
\langle J^{\mu} \rangle = \lim_{\epsilon \ra 0} \left( \frac{1}{\epsilon^4} \frac{1}{\sqrt{-\gamma}} \frac{\delta S_{reg}}{\delta A_{\mu}(\epsilon)} \right), \nonumber
\eeq
where $\g$ is the determinant of the induced metric on the $u=\epsilon$ hypersurface and $S_{reg}$ denotes the regulated action: $S_{reg} = S_{D7} + S_{CT}$. In the $B_x = 0$ case, the counterterms appearing in $S_{CT}$ were computed in ref.  \cite{O'Bannon:2007in}. A straightforward analysis reveals that no new counterterms are necessary with nonzero $B_x$ and that, as in ref. \cite{O'Bannon:2007in}, the counterterms do not contribute to $\langle J^{\mu} \rangle$. Eq. (\ref{currentdefinition}) then follows for on-shell $A_{\mu}$. For more details, see the appendix of ref. \cite{O'Bannon:2007in}},
\beq
\label{currentdefinition}
\langle J^{\mu} \rangle = \frac{\delta L}{\delta A'_{\mu}}.
\eeq
Our ansatz thus allows for a nonzero $U(1)_B$ density $\langle J^t \rangle$ as well as $U(1)_B$ currents $\jx$, $\jy$, and $\jz$. Given these constants of motion, we can solve algebraically for the derivatives of the gauge field (the field strength components):
\begin{subequations}
\label{asol}
\beq
\label{atsol}
A_t'(u) = - \frac{\sqrt{g_{uu} |g_{tt}|}}{g_{xx}^2 + \Bx^2} \frac{\jt \xi - B_z a_1}{\sqrt{\xi \chi - \frac{a_1^2}{g_{xx}^2 + \Bx^2} + \frac{a_2^2}{\gt \gx - \E^2}}},
\eeq
\beq
\label{axsol}
A_x'(u) = \sqrt{\frac{g_{uu}}{|g_{tt}|}} \frac{1}{\gx} \frac{\jx \xi - B_x a_2}{\sqrt{\xi \chi - \frac{a_1^2}{g_{xx}^2 + \Bx^2} + \frac{a_2^2}{\gt \gx - \E^2}}},
\eeq
\beq
\label{aysol}
A_y'(u) = \sqrt{\frac{g_{uu}}{|g_{tt}|}} \frac{1}{\gx} \frac{\jy \xi + E a_1}{\sqrt{\xi \chi - \frac{a_1^2}{g_{xx}^2 + \Bx^2} + \frac{a_2^2}{\gt \gx - \E^2}}},
\eeq
\beq
\label{azsol}
A_z'(u) = \frac{\sqrt{g_{uu}|g_{tt}|}}{\gt \gx - \E^2} \frac{\jz \xi - B_z a_2}{\sqrt{\xi \chi - \frac{a_1^2}{g_{xx}^2 + \Bx^2} + \frac{a_2^2}{\gt \gx - \E^2}}},
\eeq
\end{subequations}
where we have defined
\begin{subequations}
\label{coeffdefs}
\beq
\label{xidef}
\xi =  |g_{tt}| g^3_{xx} -\gx^2 \E^2 + \gt \gx \left( \Bx^2 + \Bz^2 \right) - \E^2 \Bx^2,
\eeq
\bea
\label{chidef}
\chi & = & \gt \gx^2 {\cal N}^2 (2 \pi \alpha')^4 \cos^6\theta - (2 \pi \alpha')^2 (\jx^2 + \jy^2) \\ & & + (2 \pi \alpha')^2 \left( \frac{\gt\gx}{\gx^2 + \Bx^2} \jt^2 - \frac{\gt \gx}{\gt \gx - \E^2} \jz^2 \right), \nonumber
\eea
\beq
\label{a1def}
a_1 = \pa^2 \left( \gt \gx B_z \jt + \left( \gx^2 + \Bx^2 \right) E \jy \right),
\eeq
\beq
\label{a2def}
a_2 = \pa^2 \left( \left( \gt \gx - \E^2 \right) B_x \jx + \gt \gx B_z \jz \right).
\eeq
\end{subequations}
Notice that $\xi$ is the value of $-det\left(g_{ab} + (2\pi\alpha') F_{ab} \right)$ in the $(t,x,y,z)$ subspace. It has a form characteristic of the (3+1)-dimensional Born-Infeld Lagrangian, $(-g - \frac{1}{2} g \F^2 - \frac{1}{4} (\F \wedge \F)^2)$.

We can obtain $\theta(u)$'s equation of motion in two ways. We can find its Euler-Lagrange equation of motion from the original D7-brane action, eq. (\ref{actionD7}), and then plug into that equation of motion the solutions for the field strengths in eq. (\ref{asol}). Equivalently, we can plug the solutions for the field strengths into the D7-brane action, eq. (\ref{actionD7}), to obtain an effective action for $\theta(u)$, perform a Legendre transform, and then find the Euler-Lagrange equation of motion. Plugging the solutions in eq. (\ref{asol}) into $S_{D7}$, we find
\beq
\label{onshellaction}
S_{D7} = - \N^2 (2 \pi \alpha')^2 \, \int du \, \cos^6 \theta \, \gx\sqrt{\gu \gt}\frac{\xi}{\sqrt{\xi \chi - \frac{a_1^2}{g_{xx}^2 + \Bx^2} + \frac{a_2^2}{\gt \gx - \E^2}}} .
\eeq
The Legendre-transformed on-shell action, $\hat{S}_{D7}$, is then
\bea
\label{actionLegendre}
\hat{S}_{D7}&=&S_{D7} -\int du \left( A_t' \frac{\delta S_{D7}}{\delta A_t'} +A_x'\frac{\delta S_{D7}}{\delta A_x'}+A_y'\frac{\delta S_{D7}}{\delta A_y'}+A_z'\frac{\delta S_{D7}}{\delta A_z'}\right)   \nonumber \\
&=&-\frac{1}{(2\pi \alpha')^2} \int du \, g_{xx}^{-1} \sqrt{\frac{g_{uu}}{|g_{tt}|}} \sqrt{\xi \chi - \frac{a_1^2}{g_{xx}^2 + \Bx^2} + \frac{a_2^2}{\gt \gx - \E^2}}.
\eea

To complete our solution, we must specify boundary conditions for the worldvolume fields, namely $\theta(u)$ and the gauge fields.

The boundary conditions for the gauge fields were discussed in refs. \cite{Kobayashi:2006sb,Karch:2008uy}. For $A_t(u)$, the geometry imposes a boundary condition upon us: the Killing vector corresponding to time translations becomes degenerate at the horizon, hence for the gauge field to remain well-defined as a one-form, we must impose $A_t(u_H)=0$. What about the other components of the gauge field? The key point is that the calculation of the next section implicitly fixes the values of these components at the horizon. In the next section we will demand that the on-shell action remains real for all $u$. For given values of $E$, $B_x$, $B_z$ and $\jt$, that only happens for particular values of $\jx$, $\jy$ and $\jz$. For those values of $\jx$, $\jy$, and $\jz$, the solutions for $A_x$, $A_y$ and $A_z$ are fixed by our solutions above, and hence we can then (working backwards) infer their values at the horizon. In other words, we will implicitly be choosing the values of $A_x$, $A_y$, and $A_z$ at the horizon to produce exactly the values of $\jx$, $\jy$ and $\jz$ such that the action remains real for all $u$. Unfortunately, our solution for $A_x(t,u)$ diverges at the horizon. The conductivity tensor does not depend on the values of the gauge fields at the horizon, so it is ``safe'' from the divergence. The stress-energy tensor does depend on the values at the horizon, but as explained in ref. \cite{Karch:2008uy}, these divergences (suitably regulated) have a sensible interpretation in the field theory as rates of energy and momentum loss, as we will discuss in section \ref{stressenergy}. For more details on the boundary conditions of the gauge fields, see appendix A of ref. \cite{Karch:2008uy}.

We now turn to the boundary conditions on $\theta(u)$. The field $\theta(u)$ is holographically dual to an operator that is given by taking $\frac{\partial}{\partial m}$ of the SYM Lagrangian. We will denote the operator as $\Om$. The operator $\Om$ is the $\N=2$ supersymmetric completion of the hypermultiplet fermions' mass operator, and includes several terms. The exact operator appears in ref. \cite{Kobayashi:2006sb}. For our purposes, just thinking of $\Om$ as the hypermultiplet mass operator will be sufficient. For a given solution $\theta(u)$, we can obtain the corresponding values of $m$ and $\Omv$ via a near-boundary asymptotic expansion (where the powers of $u$ follow simply from the equation of motion),
\beq
\theta(u) = \theta_1 u + \theta_3 u^3 + O\left( u^5 \right).
\eeq
As shown in refs. \cite{Karch:2005ms,Karch:2006bv}, we identify the mass as $m = \frac{\theta_1}{2 \pi \alpha'}$ and the expectation value as $\Omv \propto - 2 \theta_1 + \frac{1}{3} \theta_3^3$.

When $A_t(u)$ is zero, we have two topologically distinct ways to embed the D7-brane in the AdS-Schwarzschild background. The first type of embedding is a ``Minkowski embedding,'' in which the worldvolume $S^3$ shrinks as we move along the D7-brane away from $u=0$ and eventually collapses to a point at some $u=u'$ outside the horizon, $u' < u_H$. We then have the boundary conditions $\theta(u') = \frac{\pi}{2}$, such that $\cos \theta(u')=0$ and the $S^3$ has zero volume, and $\theta'(u') = \infty$, so that the D7-brane does not develop a conical singularity when the $S^3$ collapses to zero volume \cite{Karch:2006bv}. The D7-brane then does not extend past $u'$, but rather appears to end abruptly at $u'$.

The second type of embedding is a ``black hole'' embedding, in which the $S^3$ shrinks but does not collapse, and the D7-brane intersects the horizon. We can then choose the value of $\theta(u)$ at the horizon, $\theta(u_H) \in [0,\frac{\pi}{2})$, while for the derivative we must have $\theta'(u_H)=0$ for the embedding to be static.

When $A_t(u)$ is zero, a discontinuous (first order) transition between the two types of embeddings occurs as a function of $m/T$. The transition has been studied in great detail \cite{Babington:2003vm,Kirsch:2004km,Ghoroku:2005tf,Mateos:2006nu,Albash:2006ew,Karch:2006bv,Hoyos:2006gb,Mateos:2007vn}. Roughly speaking, large values of $m/T$ (above a critical value) correspond to Minkowski embeddings while small values of $m/T$ correspond to black hole embeddings.

As argued in ref. \cite{Kobayashi:2006sb}, however, when $A_t(u)$ is nonzero, only black hole embeddings are allowed, for a simple physical reason. With nonzero $A_t(u)$, the D7-brane has a worldvolume electric field pointing in the $u$ direction, $F_{tu}$. What source produces the electric field? The simplest possible source is a density $\jt$ of strings ending on the D7-brane. (Such a picture is nicely consistent with the field theory picture of a $U(1)_B$ density $\jt$.) A straightforward analysis then shows that the force the strings exert on the D7-brane is greater than the tension of the D7-brane \cite{Kobayashi:2006sb}. We thus expect the strings to pull the D7-brane into the black hole, producing a D7-brane black hole embedding with electric field lines in the $u$ direction.

As shown numerically in ref. \cite{Kobayashi:2006sb}, we then have a one-to-one map between values of $\theta(u_H)$ (the free parameter in the bulk) and $m = \frac{1}{2 \pi \alpha'} \theta_1$ (the free parameter near the boundary). In what follows we will not solve numerically for $\theta(u)$, however, we know the solution for $\theta(u)$ in two limits. The first limit is $m=0$, which corresponds to the trivial solution $\theta(u)=0$ and hence has $\theta(u_H)=0$. The second limit is $m \ra \infty$, where $\theta(u_H) \ra \frac{\pi}{2}$.

The phase diagram of our system has not been explored for all values of $T$, $m$, $\jt$, $E$, $B_z$, and $B_x$. To date, only certain regions, with some subset of the parameters nonzero, have been explored \cite{Babington:2003vm,Kirsch:2004km,Ghoroku:2005tf,Mateos:2006nu,Albash:2006ew,Karch:2006bv,Kobayashi:2006sb,Hoyos:2006gb,Filev:2007gb,Mateos:2007vn,Filev:2007qu,Ghoroku:2007re,Karch:2007br,Mateos:2007vc,Albash:2007bk,Erdmenger:2007bn,Albash:2007bq,Filev:2008xt,Erdmenger:2008yj,Faulkner:2008hm,Filev:2009xp,Mas:2009wf}. Where in the phase diagram will our results be valid? Our calculation of the conductivity will rely on the fact that the D7-brane intersects the horizon, so our results should be applicable in any region of the phase diagram whose description in supergravity is a D7-brane black hole embedding.

Crucially, however, as shown in refs. \cite{Filev:2007gb,Filev:2007qu,Albash:2007bk,Erdmenger:2007bn}, for the case with $B_z$ nonzero but $\jt$, $E$ and $B_x$ zero, an infinite number of solutions describing $m=0$ exist, and all but one are unstable. The stable solution is \textit{not} the trivial solution $\theta(u)=0$. On general grounds, we expect that the same should be true for our solution, which has nonzero $\jt$, $E$ and $B_x$. Nevertheless,  whenever we consider the zero mass limit, we will use the trivial solution as a simple example.

\section{The Conductivity Tensor}
\label{conductivity}

From eq. (\ref{xidef}), we see that $\xi$ is negative at the horizon but positive at the boundary, thus $\xi$ must change sign at some value of $u$, which we will call $u_*$. We can straightforwardly calculate $u_*$ from the equation\footnote{We actually find four solutions for $u_*^4/u_H^4$. The one we present is the only one for which $u_*^4/u_H^4$ takes physical values, between 0 and 1.} $\xi(u_*) = 0$, 
\beq
\label{ustar}
\frac{u_*^4}{u_H^4}=G -\sqrt{G^2-1},
\eeq 
with
\beq
G\equiv e^2-b_z^2-b_x^2+\sqrt{\left(e^2-b_z^2\right)^2+\left(b_x^2+1\right)\left(b_x^2+1+2\left(e^2+b_z^2\right)\right)},
\eeq
where we have introduced the dimensionless quantities
\beq
\label{ebc}
e\equiv \pi \alpha' u_H^2 \,E=\frac{E}{\frac{\pi}{2} \sqrt \lambda T^2}, \quad b_z\equiv \pi \alpha' u_H^2 \,B_z=\frac{B_z}{\frac{\pi}{2} \sqrt \lambda T^2}, \quad b_x\equiv \pi \alpha' u_H^2 \,B_x=\frac{B_x}{\frac{\pi}{2} \sqrt \lambda T^2}.
\eeq 
Later we will need $g_{xx}^2$ evaluated at $u_*$ in order to translate our result for the conductivity tensor into SYM theory quantities. Using eq. (\ref{ustar}), we find
\beq
\label{gxxsquared}
g_{xx}^2(u)|_{u=u_*}=\frac{\pi^4T^4}{2}(1+G)\equiv \pi^4T^4 {\cal F}(e,b_x,b_z),
\eeq
where in the  last step we removed a factor of $\pi^4 T^4$ and defined the rest to be ${\cal F}(e,b_x,b_z)$, which will appear in our result for the conductivity tensor. A useful limit is $e=0$, where $G=1$ and hence ${\cal F} = 1$.

Following refs. \cite{Karch:2007pd,O'Bannon:2007in}, we now focus on the on-shell action, eq. (\ref{onshellaction}), and in particular we focus on the square root in the denominator of eq. (\ref{onshellaction}), which we reproduce here for convenience,
\beq
\sqrt{\xi \chi - \frac{a_1^2}{g_{xx}^2 + \Bx^2} + \frac{a_2^2}{\gt \gx - \E^2}}, \nonumber
\eeq
and which also appears in the solutions for the field strengths $A_{\mu}'(u)$ for $\mu=t,x,y,z$, eq. (\ref{asol}), as well as the Legendre-transform of the on-shell action, eq. (\ref{actionLegendre}). We will argue that the four functions $\xi$, $\chi$, $a_1$ and $a_2$, must all vanish at $u_*$ in order for the above square root, and hence the on-shell action, to remain real for all $u$.

When $\xi=0$ the $a_2^2$ term is negative, because the equation $\xi(u_*) = 0$ itself tells us that $\left( \gt \gx - \E^2 \right) = - \frac{\gt \gx \Bz^2}{\left (\gx^2 + \Bx^2 \right)}< 0$ at $u_*$. To avoid an imaginary action at $u_*$ we must have $a_1(u_*)=a_2(u_*)=0$.

Arguing why $\chi$ has to vanish at $u_*$ is more subtle. $\chi$ has the same behavior as $\xi$: it is positive at the boundary and negative at the horizon, so it must have a zero at some $u$ value, which we will call $u_{\chi}$. If $u_*$ and $u_{\chi}$ are not the same, so that $\xi$ and $\chi$ have distinct zeroes, then the product $\xi \chi$ will be negative on the interval between $u_*$ and $u_{\chi}$. The crucial question then is whether the $a_2^2$ term is positive or negative on that interval. If it is positive (and sufficiently large) it could keep the action real. The sign of the $a_2^2$-term is determined by $\left( \gt \gx - \E^2 \right)$, which (like $\xi$ and $\chi$) is positive at the boundary and negative at the horizon, and hence must have have a zero at some value of $u$ that we will call $u_{E^2}$. We showed above that $\left( \gt \gx - \E^2 \right)$ is negative at $u_*$, so the zero must obey $u_{E^2} < u_*$ (it is closer to the boundary than $u_*$). Now suppose $\chi$ changes sign at $u_{\chi} >u_*$. As we just showed, the $a_2^2$ term is negative there, so the on-shell action would be imaginary on the interval ($u_*,u_{\chi}$), hence we demand $u_{\chi} \leq u_*$. We want to exclude the possibility that $u_{\chi} < u_*$. We know that $u_{E^2}$ is also less than $u_*$, so we must compare $u_{\chi}$ and $u_{E^2}$. If $u_{\chi}<u_{E^2}$, then the on-shell action is imaginary on the interval $(u_{E^2},u_*)$, and if $u_{\chi}>u_{E^2}$, the action is imaginary on the interval $(u_{\chi},u_*)$. In order for the on-shell action to remain real for all $u$, then, we demand that $u_{\chi} = u_*$.

The upshot is that we obtain four equations, $\xi(u_*) = \chi(u_*)= a_1(u_*)=a_2(u_*)=0$, for four unknows, $u_*$, $\jx$, $\jy$, and $\jz$. The equation $\xi(u_*)=0$ gives us $u_*$, as we explained above. We will now solve for the currents $\jx$, $\jy$, and $\jz$.

The equation $a_1(u_*)=0$ gives us $\jy$, while the equation $a_2(u_*) = 0$ gives us $\jz$. We then plug the results for $\jy$ and $\jz$ into $\chi(u_*)=0$ to find $\jx$. The result for the current in each case includes an overall factor of $E$, so invoking Ohm's law $\langle J^i \rangle = \sigma_{ix} \, E$, we identify the components of the conductivity tensor:
\begin{subequations}
\beq
\label{sxsol}
\sxx = \frac{\gx^2 + \Bx^2}{\gx \left(\gx^2 + \Bx^2 + \Bz^2\right)} \sqrt{\N^2 \pa^4 \gx \left( \gx^2 + \Bx^2 + \Bz^2\right) \cos^6\theta(u_*) + \pa^2 \jt^2} 
\eeq
\beq
\label{sysol}
\sxy = \frac{\pa \Bz \jt}{\gx^2 + \Bx^2 + \Bz^2}
\eeq
\beq
\label{szsol}
\sxz =\frac{\Bx \Bz}{\gx^2 + \Bx^2} \, \sxx
\eeq
\end{subequations}
where all functions of $u$ are evaluated at $u_*$. In analogy with eq. (\ref{ebc}), we define
\beq
\rho \equiv \pi \alpha' u_H^2 \,\jt = \frac{\jt}{\frac{\pi}{2} \sqrt \lambda T^2}.
\eeq
We then use the result for $g_{xx}^2(u_*)$ in eq. (\ref{gxxsquared}) to write the components of the conductivity tensor in terms of SYM theory quantities
\begin{subequations}
\label{solfield}
\beq
\label{sxsolfield}
\sxx = \sqrt{\frac{N_f^2N_c^2T^2}{16\pi^2}\frac{({\cal F} +b_x^2)^2}{\sqrt{\cal F}({\cal F} +b_x^2+b_z^2)} \cos^6\theta(u_*) + \frac{\rho^2({\cal F} +b_x^2)^2}{{\cal F}({\cal F} +b_x^2+b_z^2)^2}}
\eeq
\beq
\label{sysolfield}
\sxy = \frac{\rho \, b_z}{{\cal F} +b_x^2+b_z^2}
\eeq
\beq
\label{szsolfield}
\sxz = \frac{b_x\,b_z}{{\cal F} +b_x^2} \, \sxx \,.
\eeq
\end{subequations}

As in refs. \cite{Karch:2007pd,O'Bannon:2007in}, the result for $\sxx$ includes two terms adding in quadrature. As discussed in refs. \cite{Karch:2007pd,O'Bannon:2007in,Karch:2008uy}, these two terms have different physical interpretations. The system has two types of charge carriers. First we have the density of charge carriers we introduced explicitly in $\jt$, whose contribution appears as the second term under the square root in $\sxx$. Even when $\jt = 0$ we find a nonzero $\sxx$ and hence a nonzero current, however, so the system must have some other source of charge carriers.

The other type of charge carriers come from pair production in the electric field. Their contribution appears as the term in $\sxx$ with the $\cos^6\theta(u_*)$ factor. We have two pieces of evidence that suggests the $\cos^6\theta(u_*)$ term represents pair production. First is the behavior of the pair-production term as a function of the mass $m$. When $m \ra \infty$, so that the pair production should be suppressed, we indeed have $\cos^6\theta(u_*) \ra 0$, while when $m \ra 0$, so that the pair production should be maximal, we have $\cos^6\theta(u_*) \ra 1$. Second, as shown in ref. \cite{Karch:2008uy} for the case with $B_x=0$, when the density $\jt=0$ the flavor fields have zero momentum in the $\hat{x}$ direction, which is consistent with pair production: the oppositely-charged particles in each pair move in opposite directions, producing a finite $\jx$ but zero net momentum. For our case, with $B_x\neq0$, we see that $\s_{xz} \propto \s_{xx}$, so both types of charge carriers contribute to $\jz$, too. Using our results for the stress-energy tensor in section \ref{stressenergy}, in particular for $\langle T^t_{~x} \rangle$ and $\langle T^t_{~z} \rangle$, we can show that when $\jt=0$, the flavor fields have zero momentum in the $\hat{x}$ and $\hat{z}$ directions, so we again find a nicely consistent picture.

We will now check our result in two limits.

First, as a mild check, we set $B_x=0$, which reproduces the result of ref. \cite{O'Bannon:2007in}, in which $\vec{E}$ and $\vec{B}$ were perpendicular.

Second, following refs. \cite{Karch:2007pd,O'Bannon:2007in}, we can also take a limit of large mass. More specifically, we take $m$ to be much larger than any other scale in the problem, which includes not only $T$ but also the scale of thermal corrections to the energy of a heavy quark, $\frac{1}{2} \sqrt{\lambda} T$ \cite{Herzog:2006gh}. We will call this the ``$m \ra \infty$'' limit. As explained in section \ref{d7solution}, in that limit, $\theta(u) \ra \frac{\pi}{2}$ and hence $\cos^6 \theta(u_*) \ra 0$.

In this limit, we expect the charge carriers to behave as classical quasi-particles experiencing a drag force due to the $\N=4$ SYM plasma and a Lorentz force due to the external electric and magnetic fields. Our answer for the conductivity should then reduce to the Drude form. Let us briefly review what the Drude result is. Consider a density $\jt$ of massive quasi-particles propagating through an isotropic, homogeneous, dissipative neutral medium. In the rest frame of the medium we introduce an electric field $\vec{E}$ in the $\hat{x}$ direction, and a magnetic field $\vec{B}$ with a component $B_z$ in the $\hat{z}$ direction and a component $B_x$ in the $\hat{x}$ direction. The force on a quasi-particle is then
\beq
\label{lorentzforcelaw}
\frac{d \vec{p}}{dt} = \vec{E} + \vec{v} \times \vec{B} - \m \vec{p},
\eeq
where our quasi-particle has charge $+1$ and $\m$ is a drag coefficient. We replace the momentum with the velocity using $\vec{p} = M \vec{v}$ for quasi-particle mass $M$. We then replace the velocity with the induced current using $\vec{v} = \< \vec{J} \>/\< J^t \>$. Imposing the steady-state condition $\frac{d \vec{p}}{dt} = 0$ and solving for $\< \vec{J}\>$ yields
\beq
\sxx = \s_0 \, \frac{(B_x / \m M)^2 + 1}{|\vec{B}|^2 / (\m M)^2 + 1}, \qquad \sxy = \s_0 \, \frac{(B_z / \m M)}{|\vec{B}|^2 / (\m M)^2 + 1}, \qquad \sxz = \s_0 \, \frac{(B_x / \m M)(B_z / \m M)}{|\vec{B}|^2 / (\m M)^2 + 1},
\label{drude}
\eeq
where $\s_0 = \< J^t \> / \m M$ is the conductivity when $\vec{B}=0$.

To show that our answer reduces to the Drude result, eq. (\ref{drude}), when $m \ra \infty$, we need to know what $\mu M$ is for our charge carriers, that is, we must compute the drag force on the charge carriers, following refs. \cite{Karch:2007pd,O'Bannon:2007in}. We begin by rewriting the force law eq. (\ref{lorentzforcelaw}), in the steady state, as  
\bea
\label{drageq}
\m |\vec{p}| & = & \sqrt{E^2 + |\vec{v} \times \vec{B}|^2+ 2 \vec{E}\cdot (\vec{v} \times \vec{B})} \nonumber \\ & = & \sqrt{E^2 + v_y^2 (B_x^2+ B_z^2)+(v_z B_x-v_x B_z)^2+ 2 E_x v_y B_z} \, .
\eea
As $m\ra\infty$, pair creation will be suppressed and only the charge carriers in $\langle J^t \rangle$ will contribute to $\langle \vec{J} \rangle$, hence we may write $\langle \vec{J} \rangle = \langle J^t \rangle \vec{v}$, where we drop the $\cos \theta(u_*)$ terms in $\jx$ and $\jz$, as these vanish in our $m \ra \infty$ limit. Notice that all components of the conductivity tensor are then proportional to $\jt$, so from our answer for the conductivity tensor we find the components of $\vec{v} = \langle \vec{J} \rangle / \jt$ as functions of $E$, $B_x$ and $B_z$. What is more instructive, however, is to use the original equations $\xi(u_*)=\chi(u_*)=a_1(u_*)=a_2(u_*)=0$ to write $\vec{v}$ in terms of $g_{xx}(u_*)$ and $g_{tt}(u_*)$. For example, the speed of the heavy charge carriers is
\beq
|\vec{v}| = \left . \sqrt{\frac{|g_{tt}|}{g_{xx}}} \right|_{u_*},
\eeq
which is the local speed of light at $u_*$. The drag force is
\beq
\m |\vec{p}| = \frac{1}{2\p\a'} \sqrt{|g_{tt}(u_*)|g_{xx}(u_*) },
\label{dragforce}
\eeq
which is simply the Nambu-Goto Lagrangian (density) for a string extended in the $\hat{x}$ direction, sitting at fixed radial position $u_*$. Following refs. \cite{Herzog:2006gh,Gubser:2006bz,Karch:2007pd,O'Bannon:2007in}, if we employ the relativistic relation $|\vec{p}| = \gamma M v$ with $\gamma = \frac{1}{\sqrt{1-v^2}}$ and $M$ the quasi-particle mass, then we find
\beq
\m M = \frac{1}{2\p\a'} \sqrt{g_{xx}(u_*)^2-|g_{tt}(u_*)|g_{xx}(u_*)}= \frac{\p}{2} \sqrt{\lam} T^2 \ ,
\label{mum}
\eeq
which is identical to the zero-density result of refs. \cite{Herzog:2006gh,Gubser:2006bz} and the finite density results of refs. \cite{Karch:2007pd,O'Bannon:2007in}, but now with nonzero $B_x$. That we recover the same answer is not surprising in the probe limit $N_f \ll N_c$. In the probe limit, the flavor excitations are too dilute to experience a significant number of collisions with one another. Most of their energy loss comes from their interactions with the $\N=4$ SYM plasma, rather than with other flavor excitations, hence the drag force is independent of $\jt$. See refs. \cite{O'Bannon:2007in,O'Bannon:2008aj} for more detailed explanations.

We can now compare to the Drude form eq. (\ref{drude}). We take $m \ra \infty$, so that $\cos^6\theta(u_*) \ra 0$ in the conductivity tensor. We also ``linearize'' in the electric field, that is, we consider the regime of linear response, where the currents are linear in $E$ and hence the conductivity is constant in $E$. (Recall that the Drude form relies on Maxwell's equations, which are linear.) In practical terms, that means setting $E=0$ in our result for the conductivity. That means we take ${\cal F} (e=0,b_x,b_z)=1$ as explained above. Lastly, using our identification of $\mu M$ in eq. (\ref{mum}), we can write
\beq
\rho = \frac{\jt}{\frac{\pi}{2} \sqrt{\lambda} T^2} = \frac{\jt}{\mu M},
\eeq
and similarly for $b_x$ and $b_z$ (recall eq. (\ref{ebc})). We immediately find that our result for the conductivity tensor is identical to the Drude form, eq. (\ref{drude}).

Finally, given that the novelty of our result is the presence of $B_x$, we can take limits that highlight the effects of $B_x$. For example, we can show that, generically, $B_x$ enhances the process of pair production. We first linearize in the electric field again, so ${\cal{F}} = 1$, and then isolate the pair-production term by taking zero density ($\jt = 0$, hence $\rho=0$). The result for $\s_{xx}$ is then
\beq
\s_{xx} = \frac{N_f N_c T}{4\pi} \frac{1+b_x^2}{\sqrt{1+b_x^2 + b_z^2}} \cos^3\theta(u_*).
\eeq
If we further consider $b_x \gg b_z$, then we see that $\s_{xx}$ has a $\sqrt{1+b_x^2}$ factor. Clearly, increasing $B_x$ increases the contribution to $\jx$ from pair production. Conversely, if we suppress the pair production by taking $m \ra \infty$, so that $\cos^6\theta(u_*) \ra 0$, while keeping $\jt$ finite, then $\s_{xx}$ reduces to
\beq
\s_{xx} =  \rho \, \frac{1+b_x^2}{1+b_x^2 + b_z^2},
\eeq
(which is of course the Drude result from eq. (\ref{drude})) so that now taking $b_x \gg b_z$ we find that $\s_{xx} \ra \rho$. Increasing $B_x$ does not enhance the contribution to $\jx$ coming from the net density $\jt$ of charge carriers. (By contrast, the limit $b_z \gg b_x$ clearly suppresses both contributions to the current.)

\section{The Stress-Energy Tensor}
\label{stressenergy}

In this section we use our holographic setup to compute the contribution that the flavor fields make to the expectation value of the stress-energy tensor of the field theory. We will call this contribution $\langle T^{\mu}_{~\nu} \rangle$. We also identify certain divergences in the stress-energy tensor which are related to the rates of energy and momentum loss of the charge carriers (the flavor fields). We also discuss two special quantities that are free from these IR divergences. This section is a direct extension of the results of ref. \cite{Karch:2008uy} to include nonzero $B_x$.

Many contributions to the stress-energy tensor come simply from the electric polarization and the magnetization of the medium, as we will now review. Even in an equilibrium system, background electric and magnetic fields produce non-vanishing momentum currents due to polarization effects, so that we expect a contribution to $\langle T_{\mu \nu} \rangle$ of the form
\beq
\langle T^{\mu}_{~\nu} \rangle_{pol} = M^{\mu}_{~ \sigma} \, F^{\sigma}_{~ \nu}.
\eeq
where $M^{\mu \nu}$ is the polarization tensor,
\beq
M^{\mu \sigma} = - \frac{\delta \Omega}{\delta F_{\mu \sigma}},
\eeq
with $\Omega$ the free energy density (and where we take the variation with other variables held fixed). The components of $M^{\mu \sigma}$ with one $t$ index and one spatial index are electric polarizations while components with two spatial indices are magnetizations. The full energy-momentum tensor $\langle T^{\mu}_{~\nu} \rangle$ then divides into two pieces:
\beq
\langle T^{\mu}_{~\nu} \rangle = \langle T^{\mu}_{~\nu} \rangle_{fluid} + \langle T^{\mu}_{~\nu} \rangle_{pol},
\eeq
where, for example, $\langle T^{t}_{~i} \rangle_{fluid}$ corresponds to the genuine momentum current due to the flow in the medium. Both $\langle T^{\mu}_{~\nu} \rangle$ and $\langle T^{\m}_{~\n} \rangle_{fluid}$ obey the same (non-)conservation equation,
\beq
\partial^{\mu} \langle T_{\mu \nu} \rangle = F_{\nu \rho} \langle J^{\rho} \rangle,
\eeq
but only $\langle T^{\m}_{~\n} \rangle_{fluid}$ represents observable quantities that can couple to external probes of the system (and hence is the appropriate object to use when studying transport).

In gauge-gravity duality, we identify $\Omega = - S_{D7}$, where here $S_{D7}$ is the D7-brane action evaluated on a particular solution for the worldvolume fields, so that
\beq
M^{\mu \nu} = \frac{\delta S_{D7}}{\delta F_{\mu \nu}}.
\eeq
As an explicit example, consider for example the calculation of $M^{tx}$. We start with eq. (\ref{originaldbi}), evaluated on a particular solution. The on-shell action $S_{D7}$ will have explicit $E$ dependence, as well as implicit dependence through the solutions for $\theta(u)$ and the worldvolume gauge fields. We thus employ the chain rule\footnote{We are using arguments similar to those in refs. \cite{Mateos:2007vn,Albash:2007bk,O'Bannon:2008bz}.},
\beq
\frac{d S_{D7}}{d E} = - \int du \, \left [ \frac{\partial L}{\partial E} \, + \, \frac{\partial \theta}{\partial E} \frac{\partial L}{\partial \theta} \, + \, \frac{\partial \theta'}{\partial E} \frac{\partial L}{\partial \theta'} \, + \, \sum_{\mu=t,x,y,z} \frac{\partial A_{\mu}'}{\partial E} \frac{\partial L}{\partial A_{\mu}'} \right].
\eeq
We then use the fact that partial derivatives commute to write $\frac{\partial}{\partial E} \frac{\partial}{\partial u} = \frac{\partial}{\partial u} \frac{\partial}{\partial E}$, and integrate by parts to find
\bea
\label{bulkpolarizationequation}
\frac{d S_{D7}}{d E} & = & - \int du \, \left [ \frac{\partial L}{\partial E} \, + \, \left( \frac{\partial L}{\partial \theta} \, - \, \frac{\partial}{\partial u} \frac{\partial L}{\partial \theta'} \right) \frac{\partial \theta}{\partial E} \, - \, \sum_{\mu=t,x,y,z} \frac{\partial A_{\mu}}{\partial E} \frac{\partial}{\partial u} \frac{\partial L}{\partial A_{\mu}'} \right] \nonumber \\ & & \,\,\,\,\,\,\, - \left . \frac{\partial \theta}{\partial E} \frac{\partial L}{\partial \theta'} \right |_{0}^{u_H} - \left . \sum_{\mu=t,x,y,z} \frac{\partial A_{\mu}}{\partial E} \frac{\partial L}{\partial A_{\mu}'} \right |_{0}^{u_H}.
\eea
Of the terms under the integral, the term in parentheses and the terms in the sum over $\mu$ vanish due to the equations of motion. That leaves the $\frac{\partial L}{\partial E}$ term under the integral, and the boundary terms. The main point is that the only contribution to the polarization from the bulk of $AdS_5$ comes from $\frac{\partial L}{\partial E}$. Similar statements apply for the magnetizations, for example, for $M^{xy}$ the only bulk term comes from $\frac{\partial L}{\partial B_z}$.

In fact, we find that all six components of the polarization tensor are nonzero. All three electric polarizations, $M^{ti}$ with $i=x,y,z$, are nonzero, despite the fact that our solution describes an electric field only in the $\hat{x}$ direction. In other words, if, for example, we introduce an electric field in the $\hat{y}$ direction, $E_y$, take the variation of $S_{D7}$ with respect to $E_y$, and then set $E_y = 0$, we find a nonzero answer. Similarly, all three components of the magnetization are nonzero although our solution only includes $B_x$ and $B_z$. In all cases the only bulk contribution is from a $\frac{\partial L}{\partial F_{\mu \nu}}$ term, evaluated on our solution (where only $E_x$, $B_x$ and $B_z$ are nonzero). We present explicit expressions for the derivatives $\frac{\partial L}{\partial F_{\mu \nu}}$ in the appendix. We will shortly see the derivatives $\frac{\partial L}{\partial F_{\mu \nu}}$ appearing in the stress-energy tensor. Most of these arise from the expected contribution to $\langle T_{\mu \nu} \rangle$ from $\langle T^{\mu}_{~\nu} \rangle_{pol}$.

We now come to the calculation of the stress-energy tensor. As explained in ref. \cite{Karch:2008uy}, we may invoke the Hamiltonian form of the AdS/CFT correspondence \cite{Witten:1998qj}, which allows us to equate conserved charges in the boundary field theory and the bulk gravity theory. For example, if $p_i$ denotes the momentum associated with the flavor fields in the SYM theory, with $i = x,y,z$, then in the Hamiltonian framework we identify the conserved charges
\beq
p_i = \int dt \, d\vec{x} ~\langle T^t_{~i} \rangle = \int dt \, d\vec{x} \,du \, d^3\alpha ~\sqrt{-g_{D7}} \, \, \Ui.
\eeq
The $\alpha$ are coordinates on the $S^3$ wrapped by the D7-branes, $g_{D7}$ is the determinant of the induced metric on the D7-branes, and $\Ui$ is the D7-branes' momentum density. If the energy-momentum tensors are independent of the four spacetime coordinates, then the integrals over $dt \, d\vec{x}$ will only produce a factor of the spacetime volume, so that we can equate the momentum densities directly:
\beq
\label{FTT}
\Ti = \int du \, d^3\alpha ~\sqrt{-g_{D7}} \, \, \Ui.
\eeq
To compute the stress-energy tensor of the flavor fields, then, we must compute the stress-energy tensor of the D7-branes, $\Theta^a_{~b}$, defined as
\beq
\label{d7setensor}
\Theta^a_{~b} \equiv \int \, du \, d^3\alpha ~\sqrt{-g_{D7}} \, U^{a}_{~b}.
\eeq
When the indices $a$ and $b$ are in SYM theory directions, we can identify $\langle T^a_{~b}\rangle = \Theta^a_{~b}$. The indices $a$ and $b$ can also be in the $u$ or $S^3$ directions, however, in which case the SYM theory interpretation is more difficult. Following ref. \cite{Karch:2008uy}, we will be able to provide a field theory interpretation for some, but not all, components.

We can compute $\Theta^a_{~b}$ in two different ways. One way is to compute the variation of the D7-brane action, $S_{D7}$, with respect to the background metric. The other way is to use a Noether procedure, since the momenta are the generators of translation symmetries. We have used both methods and have found perfect agreement. The calculation by variation of the action is longer and more difficult than the Noether procedure, however, so we will not present it. The result of the Noether procedure is
\beq
\label{noetherstressenergy}
\Theta^a_{~b} = -\int du \, \left( L \, \delta^{a}_{~b} + 2 F_{c b} \frac{ \delta L }{\delta F_{a c} } - \partial_{b}\theta \, \frac{\delta L}{\delta \partial_a \theta} \right),
\eeq
where we have performed the trivial integration over the $S^3$.

We expect the last term in eq. (\ref{noetherstressenergy}) to contribute to $T^{u}_{~u}$, given our ansatz $\theta(u)$. We find, however, that the last term in eq. (\ref{noetherstressenergy}) also contributes to the $T^{\mu}_{~u}$ components with $\mu = t,x,y,z$. In other words, suppose we allow $\theta$ to depend on $t,x,y,z$. We then find that, taking the derivatives $\frac{\delta L}{\delta \partial_{\mu} \theta}$, with $\mu=t,x,y,z$, and then setting $\partial_{\mu} \theta = 0$ produces a nonzero result. This is very similar to what we saw for the polarization tensor above, where all six components were nonzero even though our solution has only $E$, $B_x$ and $B_z$ nonzero. We write explicit expressions for the derivatives $\frac{\delta L}{\delta \partial_{\mu} \theta}$ in the appendix.

We will now present all the components of the stress-energy tensor.

In the $S^3$ directions the only components are on the diagonal, and all are simply $-\int du \, L = S_{D7}$. The nontrivial components are in the $(u,t,x,y,z)$ subspace. For notational simplicity, we will identify current components, $\langle J^{\mu} \rangle$, whenever possible, and we will not write $\int du$, which appears for every component. Primes denote $\frac{\partial}{\partial u}$.

The components with upper index $t$ are
\beq
\begin{array}{ccccc}
\Theta^t_{~t} & = & -L - F_{xt} \frac{\delta L}{\delta F_{tx}} - F_{ut} \frac{\delta L}{\delta F_{tu}} & = & -L + E_x \frac{\partial L}{\partial E_x} + \langle J^t \rangle A_t' \bigskip \\
\Theta^t_{~x} & = & - F_{ux} \frac{\delta L}{\delta F_{tu}} - F_{yx} \frac{\delta L}{\delta F_{ty}} & = &  \langle J^t \rangle A_x' -\frac{\partial L}{\partial E_y} B_z \bigskip \\
\Theta^t_{~y} & = & -F_{xy} \frac{\delta L}{\delta F_{tx}} - F_{zy} \frac{\delta L}{\delta F_{tz}} - F_{uy} \frac{\delta L}{\delta F_{tu}} & = &  B_z \frac{\partial L}{\partial E_x} - B_x \frac{\partial L}{\partial E_z} + \langle J^t \rangle A_y' \bigskip \\
\Theta^t_{~z} & = & -F_{yz} \frac{\delta L}{\delta F_{ty}} - F_{uz} \frac{\delta L}{\delta F_{tu}} & = &  B_x \frac{\partial L}{\partial E_y} + \langle J^t \rangle A_z' \bigskip \\
\Theta^t_{~u} & = & -F_{xu} \frac{\delta L}{\delta F_{tx}} - F_{yu} \frac{\delta L}{\delta F_{ty}} - F_{zu} \frac{\delta L}{\delta F_{tz}} +\theta' \frac{\delta L}{\delta \partial_t \theta}& = &- A_x' \frac{\partial L}{\partial E_x} - A_y' \frac{\partial L}{\partial E_y} - A_z' \frac{\partial L}{\partial E_z} +\theta' \frac{\delta L}{\delta \partial_t \theta}\nonumber
\end{array}
\eeq

The components with upper index $x$ are
\beq
\begin{array}{ccccc}
\Theta^x_{~t} & = & -F_{ut} \frac{\delta L}{\delta F_{xu}} & = &  \jx A_t' \bigskip \\
\Theta^x_{~x} & = &- L - F_{tx} \frac{\delta L}{\delta F_{xt}} - F_{yx} \frac{\delta L}{\delta F_{xy}} - F_{ux} \frac{\delta L}{\delta F_{xu}} & = & - L + E_x \frac{\partial L}{\partial E_x} + B_z \frac{\partial L}{\partial B_z} + \jx A_x' \bigskip \\
\Theta^x_{~y} & = & - F_{zy} \frac{\delta L}{\delta F_{xz}} - F_{uy} \frac{\delta L}{\delta F_{xu}} & = & -B_x \frac{\partial L}{\partial B_y} +  \jx A_y' \bigskip \\
\Theta^x_{~z} & = & - F_{yz} \frac{\delta L}{\delta F_{xy}} - F_{uz} \frac{\delta L}{\delta F_{xu}} & = & -B_x \frac{\partial L}{\partial B_z} +  \jx A_z' \bigskip \\
\Theta^x_{~u} & = &- F_{tu} \frac{\delta L}{\delta F_{xt}} - F_{yu} \frac{\delta L}{\delta F_{xy}} - F_{zu} \frac{\delta L}{\delta F_{xz}} + \theta' \frac{\delta L}{\delta \partial_x \theta} & = &  A_t' \frac{\partial L}{\partial E_x} + A_y' \frac{\partial L}{\partial B_z} - A_z' \frac{\partial L}{\partial B_y} + \theta' \frac{\delta L}{\delta \partial_x \theta} \nonumber
\end{array}
\eeq

The components with upper index $y$ are
\beq
\begin{array}{ccccc}
\Theta^y_{~t} & = & - F_{xt} \frac{\delta L}{\delta F_{yx}} - F_{ut} \frac{\delta L}{\delta F_{yu}} & = &  E_x \frac{\partial L}{\partial B_z} + \jy A_t' \bigskip \\
\Theta^y_{~x} & = & -F_{ux} \frac{\delta L}{\delta F_{yu}} - F_{tx} \frac{\delta L}{\delta F_{yt}} & = &  E_x \frac{\partial L}{\partial E_y} + \jy A_x' \bigskip \\
\Theta^y_{~y} & = & -L - F_{xy} \frac{\delta L}{\delta F_{yx}} - F_{zy} \frac{\delta L}{\delta F_{yz}} - F_{uy} \frac{\delta L}{\delta F_{yu}} & = &- L + B_z \frac{\partial L}{\partial B_z} + B_x \frac{\partial L}{\partial B_x} + \jy A_y' \bigskip \\
\Theta^y_{~z} & = & - F_{uz} \frac{\delta L}{\delta F_{yu}} & = & \jy A_z' \bigskip \\
\Theta^y_{~u} & = &- F_{tu} \frac{\delta L}{\delta F_{yt}} - F_{xu} \frac{\delta L}{\delta F_{yx}} - F_{zu} \frac{\delta L}{\delta F_{yz}} + \theta' \frac{\delta L}{\delta \partial_y \theta} & = &  A_t' \frac{\partial L}{\partial E_y} - A_x' \frac{\partial L}{\partial B_z} + A_z' \frac{\partial L}{\partial B_x} + \theta' \frac{\delta L}{\delta \partial_y \theta} \nonumber
\end{array}
\eeq

The components with upper index $z$ are
\beq
\begin{array}{ccccc}
\Theta^z_{~t} & = & - F_{xt} \frac{\delta L}{\delta F_{zx}} - F_{ut} \frac{\delta L}{\delta F_{zu}} & = & - E_x \frac{\partial L}{\partial B_y} + \jz A_t' \bigskip \\
\Theta^z_{~x} & = & -F_{ux} \frac{\delta L}{\delta F_{zu}} - F_{tx} \frac{\delta L}{\delta F_{zt}} - F_{yx} \frac{\delta L}{\delta F_{zy}} & = &  E_x \frac{\partial L}{\partial E_z} - B_z \frac{\partial L}{\partial B_x} + \jz A_x' \bigskip \\
\Theta^z_{~y} & = & - F_{xy} \frac{\delta L}{\delta F_{zx}} - F_{uy} \frac{\delta L}{\delta F_{zu}} & = & - B_z \frac{\partial L}{\partial B_y} + \jz A_y' \bigskip \\
\Theta^z_{~z} & = & -L - F_{yz} \frac{\delta L}{\delta F_{zy}} - F_{uz} \frac{\delta L}{\delta F_{zu}} & = & -L + B_x \frac{\partial L}{\partial B_x} + \jz A_z' \bigskip \\
\Theta^z_{~u} & = &- F_{tu} \frac{\delta L}{\delta F_{zt}} - F_{xu} \frac{\delta L}{\delta F_{zx}} - F_{yu} \frac{\delta L}{\delta F_{zy}} + \theta' \frac{\delta L}{\delta \partial_z \theta} & = &  A_t' \frac{\partial L}{\partial E_z} + A_x' \frac{\partial L}{\partial B_y} - A_y' \frac{\partial L}{\partial B_x} + \theta' \frac{\delta L}{\delta \partial_z \theta} \nonumber
\end{array}
\eeq

The components with upper index $u$ are
\beq
\begin{array}{ccccc}
\Theta^u_{~t} & = & - F_{xt} \frac{\delta L}{\delta F_{ux}} & = & - \jx E_x \bigskip \\
\Theta^u_{~x} & = & - F_{tx} \frac{\delta L}{\delta F_{ut}} - F_{yx} \frac{\delta L}{\delta F_{uy}} & = &  \langle J^t \rangle E_x + \jy B_z \bigskip \\
\Theta^u_{~y} & = & -F_{xy} \frac{\delta L}{\delta F_{ux}} -F_{zy} \frac{\delta L}{\delta F_{uz}} & = & - \jx B_z + \jz B_x \bigskip \\
\Theta^u_{~z} & = & -F_{yz} \frac{\delta L}{\delta F_{uy}} & = & - \jy B_x \bigskip \\
\Theta^u_{~u} & = & -L - \sum_{\mu=t,x,y,z} F_{\mu u} \frac{\delta L}{\delta F_{u\mu}} + \theta' \frac{\delta L}{\delta \theta'} & = & - L + \sum_{\mu=t,x,y,z} \langle J^{\mu} \rangle A_{\mu}' + \theta' \frac{\delta L}{\delta \theta'} \nonumber
\end{array}
\eeq
All quantities on the right-hand sides are evaluated on-shell.

We would like to convert the components of $\Theta^a_{~b}$ to field theory quantities. In most cases, whether we can do so depends on whether we can perform the $u$ integration. Sometimes this is easy. For example, we know that $\int \, du \, L = -S_{D7} = \Omega$, and $\int \, du \, A_t'(u) = - \mu$, where $\mu$ is the $U(1)_B$ chemical potential. In some cases we can translate to SYM theory quantities without doing the $u$ integrals. For instance, terms with the derivatives $\frac{\partial L}{\partial F_{\mu \nu}}$ multiplying the $u$-independent quantities $E_x$, $B_x$, or $B_z$ we can interpret as contributions from the polarization tensor, as explained above. On the other hand, we have not found a field theory interpretation for the components $\Theta^{\mu}_{~u}$ with $\mu=t,x,y,z$ because the $u$ integration is non-trivial. For many components, converting to SYM theory quantities requires integrating $A_x'$, $A_y'$, or $A_z'$, for which the field theory meaning is not immediately clear.

As discussed in ref. \cite{Karch:2008uy} (following ref. \cite{Gubser:2008vz}), the components $\Theta^u_{~\mu}$, with $\mu=t,x,y,z$, do have a clear interpretation in the SYM theory: they are proportional to rates of energy or momentum loss. To explain this, we return to the field theory side of the correspondence. Recall that in the presence of background electric and magnetic fields, the (non-)conservation law for the stress-energy tensor was
\beq
\label{rates}
\partial^{\mu} \langle T_{\mu \nu} \rangle = F_{\nu \rho} \langle J^{\rho} \rangle.
\eeq
For our spatially homogeneous solution, all the spatial derivatives on the left-hand side will vanish, leaving only the time derivatives. With our background fields and current, we thus have
\bea
\label{ratesexplicit}
\partial_t \langle T^t_{~t} \rangle & = & - E \langle J^x \rangle \\
\partial_t \langle T^t_{~x} \rangle & = & E \langle J^t \rangle + B_z \langle J^y \rangle \nonumber \\
\partial_t \langle T^t_{~y} \rangle & = & - B_z \langle J^x \rangle + B_x \langle J^z \rangle \nonumber \\
\partial_t \langle T^t_{~z} \rangle & = & - B_x \langle J^y \rangle. \nonumber
\eea
Our system also has a net density of charge carriers in an external electric field. The electric field is thus doing net work on the system. The charge carriers (the flavor degrees of freedom) will transfer energy and momentum to the $\N=4$ SYM plasma, so that, over time, the $\N=4$ SYM plasma will heat up, and begin to move. Eq. (\ref{ratesexplicit}) tells us the rates at which the energy and momentum of the flavor degrees of freedom are changing.

The energy and momentum loss rates on the right-hand-side of eq. (\ref{ratesexplicit}) are identical to the components of the stress-energy tensor with upper index $u$ and lower index $\mu = t,x,y,z$, the $\Theta^u_{~\mu}$, up to a constant factor. In the expressions above for the $\Theta^u_{~\mu}$, the constant factor comes from the integration over $u$ (suppressed for notational clarity), which produces a factor $\int_0^{u_H} du = u_H = \frac{\sqrt{2}}{\pi T}$. The holographic calculation thus encodes the energy and momentum loss rates in the components of the stress-energy tensor with upper index $u$, as previously discussed in refs. \cite{Gubser:2008vz,Karch:2008uy}.

As an important aside, notice that our system has translation invariance, which implies momentum conservation. In other words, the system appears to have no mechanism for dissipation of momentum. Why then do we find a finite Ohmic conductivity, $\sigma_{xx}$? The answer comes from the probe limit, $N_f \ll N_c$. The very dilute flavor degrees of freedom will indeed transfer energy and momentum to the far more numerous $\N=4$ SYM degrees of freedom, but the rates at which they do so are of order $N_f N_c$, as we can see from eq. (\ref{ratesexplicit}). The rates go as factors of the $\langle J^{\mu} \rangle$ components times the external fields $E$, $B_x$ and $B_z$. The $\langle J^{\mu} \rangle$ that we study are order $N_f N_c$, while the external fields are order one in the large $N_c$ counting. We may thus conclude that only after a time on the order of $N_c$ will the flavor degrees of freedom have transferred an order $N_c^2$ amount of energy and momentum to the $\N=4$ SYM plasma. For earlier times, we may safely ignore the motion of the plasma, that is, the plasma will act as a reservoir into which the flavor fields may ``dump'' energy and momentum. For those early times, then, the probe limit allows the system to mimic a dissipative system, and hence we find our finite Ohmic conductivity. At late times (on the order of $N_c$), however, we could no longer ignore the motion of the plasma (and hence we would need to do a new calculation of the conductivity and stress-energy tensors).

Back on the supergravity side of the correspondence, the loss rates in eq. (\ref{ratesexplicit}) also appear as divergences in the corresponding components of the D7-brane's stress-energy tensor, as explained in ref. \cite{Karch:2008uy}. Specifically, the energy and momentum densities $\Theta^t_{~\mu}$ exhibit divergences coming from the $u=u_H$ endpoint of the $u$ integration (which was suppressed for notational clarity above). Such divergences are familiar from the dragging string solution of refs. \cite{Herzog:2006gh,Gubser:2006bz}, which represented a field theory process in which a single heavy charge carrier lost energy and momentum to the SYM plasma. We are thus not too surprised to see similar divergences here, where we have a density of charge carriers.

The divergences in $\Theta^t_{~\mu}$ appear to come from two sources. One is a divergence in our solution for $A_x'(u)$. If we Taylor expand our solution for $A_x'(u)$ in powers of $|g_{tt}|$, we find
\beq
\label{axdivergence}
A_x'(u) = - E \, \sqrt{\frac{g_{uu}}{|g_{tt}|}} + O\left( \sqrt{|g_{tt}|} \right),
\eeq
so that $\int du \, A_x'(u)$, which appears in $\Theta^t_{~x}$, produces a divergence at the $u=u_H$ endpoint. In contrast, the other field strengths, $A_t'(u)$, $A_y'(u)$, and $A_z'(u)$, all vanish at the horizon (the leading term in their expansions in $\sqrt{|g_{tt}|}$) and hence these produce no divergences at $u=u_H$.

The second source of divergences is from the derivatives $\frac{\partial L}{\partial E_i}$ with $i=x,y,z$. These are the bulk contributions to the electric polarizations, as explained above. Performing a Taylor expansion in $|g_{tt}|$ for these, we find
\beq
\label{polarizationdivergence}
\frac{\partial L}{\partial E_i} = \langle J^i \rangle \sqrt{\frac{g_{uu}}{|g_{tt}|}} + O\left( \sqrt{|g_{tt}|} \right).
\eeq
In the $\Theta^t_{~\mu}$, these appear multiplied by $E$, $B_x$ and $B_z$, so that the integral over $u$ produces a divergence at $u=u_H$. We note in passing that $\frac{\partial L}{\partial B_x}$, $\frac{\partial L}{\partial B_y}$, and $\frac{\partial L}{\partial B_z}$ have no such divergences (for each, the leading term is $\sqrt{|g_{tt}|}$).

Following ref. \cite{Karch:2008uy}, we can relate the \textit{coefficients} of the divergent terms with the loss rates in eq. (\ref{ratesexplicit}) as follows. On the SYM theory side, the divergences comes from the infra-red (IR): the charges have been losing energy and momentum at constant rates for infinite \textit{time}. To regulate the divergence, then, we want to consider charges moving for some finite time $\Delta t$. We can then identify the divergences in the $\Theta^t_{~\mu}$ as the constant rates times $\Delta t$: $\partial_t \langle T^t_{~\mu} \rangle \Delta t$. On the supergravity side, we should only include those parts of the spacetime that had time to communicate with the boundary in the time $\Delta t$. In particular, we would like the boundary to communicate with the horizon. We thus define $\Delta t$ as the time required for a light ray to travel from the boundary to the horizon,
\beq
\label{deltat}
\Delta t = \int_0^{u_H - \epsilon} du \, \sqrt{\frac{g_{uu}}{|g_{tt}|}},
\eeq
where we have introduced a regulator to make $\Delta t$ finite: we integrate not to the horizon $u_H$ but to some $u_H - \epsilon$. As $\epsilon \ra 0$, $\Delta t$ diverges as $\frac{1}{\epsilon}$. Clearly the divergences in the $\Theta^t_{~\mu}$ are of the form in eq. (\ref{deltat}). We thus plug eqs. (\ref{axdivergence}) and (\ref{polarizationdivergence}) into our expressions for the $\Theta^t_{~\mu}$ above, and using eq. (\ref{deltat}), we immediately reproduce the right-hand side of eq. (\ref{ratesexplicit}). The holographic calculation thus encodes the energy and momentum loss rates in the coefficients of the $u=u_H$ divergences of the $\Theta^t_{~\mu}$, as discussed previously in ref. \cite{Karch:2008uy}.

As also discussed in ref. \cite{Karch:2008uy}, we can find observers who will not see the charges lose any energy or momentum. These observers will thus see no divergences; the energy and momenta they measure will be ``IR-safe.'' We will identify two such observers, who we will call observer 1 and observer 2.

Observer 1 moves along with the charges. In that observer's reference frame, the charges are at rest (and the surrounding plasma is moving past), so obviously observer 1 will not see the charges lose energy or momentum. Observer 1 should thus see no divergences. More formally, observer 1 will have a four-velocity proportional to the charge current, $v_1^{\m} \propto \langle J^{\m} \rangle$. Notice that $v_1^{\mu}$ is thus covariantly constant, $\partial_{\mu} v_1^{\nu} = 0$. The mass-energy four-vector associated with observer 1 is then proportional to
\beq
I_1^{\mu} = \langle T^{\mu}_{~\nu} \rangle \, v_1^{\nu} \propto \langle T^{\mu}_{~\nu} \rangle \, \langle J^{\nu} \rangle, \nonumber
\eeq
and using $\partial_{\mu} \langle T^{\mu}_{~\nu} \rangle = F_{\nu \alpha} \langle J^{\alpha} \rangle$, we can easily show that $\partial_{\mu} I_1^{\mu} = F_{\alpha \beta} \langle J^{\alpha} \rangle \langle J^{\beta} \rangle= 0$, so $I_1^{\mu}$ is a conserved current. Furthermore, $I_1^{\mu}$ is free of divergences, that is, if we write the $t$ component explicitly,
\beq
I^t_1 = \langle T^t_{~t} \rangle \jt + \langle T^t_{~x} \rangle \jx + \langle T^t_{~y} \rangle \jy + \langle T^t_{~z} \rangle \jz,
\eeq
and insert our expressions for the $\langle T^t_{~\mu} \rangle$ from our $\Theta^t_{~\mu}$, we find that all the divergences (of the form $\sqrt{g_{uu}/|g_{tt}|}$) cancel exactly.

Observer $2$ has four-velocity $v_2^{\mu} \propto \epsilon^{\mu \alpha \beta \gamma} F_{\alpha \beta} \langle J_{\gamma} \rangle$. Notice that $v_2^{\mu}$ is again covariantly constant, $\partial_{\mu} v_2^{\nu} = 0$, because the currents and external fields are constant. Notice also that observer $2$ is moving orthogonally to observer $1$, that is, their four-velocities are orthogonal: $v_1^{\mu} v_{2 \, \mu} \propto \langle J_{\mu} \rangle\, \epsilon^{\mu \alpha \beta \gamma} \, F_{\alpha \beta} \, \langle J_{\gamma} \rangle = 0$. In fact, in the language of section 4.2 of ref. \cite{Wald:1984rg}, $v_2^{\mu}$ is (proportional to) the magnetic field as measured by observer 1. The mass-energy four-vector of observer $2$ is
\beq
I_2^{\mu} = \langle T^{\mu}_{~\nu} \rangle \, v_2^{\nu} \propto \langle T^{\mu}_{~\nu} \rangle \, \epsilon^{\nu \alpha \beta \gamma} \, F_{\alpha \beta} \, \langle J_{\gamma} \rangle, \nonumber
\eeq
and again using $\partial_{\mu} \langle T^{\mu}_{~\nu} \rangle = F_{\nu \alpha} \langle J^{\alpha} \rangle$, we can easily show that $\partial_{\mu} I_2^{\mu} \propto \left( F \wedge F \right) \, J^2$, where $J^2 = J^{\mu} \, J_{\mu}$, so $I_2^{\mu}$ is only a conserved current when $F \wedge F \propto \vec{E} \cdot \vec{B} = 0$. In other words, when $\vec{E} \cdot \vec{B}$ is nonzero we should have $\partial_t I^t_2 \propto \left( F \wedge F\right)J^2$, so that, as we saw for the stress-energy tensor, we should find a divergence in $I^t_2$ whose coefficient is the loss rate, $\left(F \wedge F\right) J^2$. Indeed, a straightforward calculation shows that $I^t_2$ includes the usual $\sqrt{g_{uu}/|g_{tt}|}$ divergence, with coefficient $\left( F \wedge F\right) J^2$. Observer 2 only sees an ``IR-safe'' conserved current $I_2^{\mu}$ when $\vec{E} \cdot \vec{B} = 0$.

\section{Conclusion}
\label{conclusion}

Using the holographic setup described in section \ref{d7solution}, we computed the conductivity tensor, and the contribution to the stress-energy tensor, of $\N=2$ supersymmetric flavor fields propagating through a strongly-coupled $\N=4$ SYM theory plasma at temperature $T$. We included a finite $U(1)_B$ density $\jt$ and considered the most general configuration of constant external fields, namely an electric field $E$ and a magnetic field with a component $B_z$ perpendicular to $E$ and a component $B_x$ parallel to $E$. We also discussed divergences in the flavor fields' contribution to the stress-energy tensor, and discussed some ``IR-safe'' quantities that are free from these divergences.

We will suggest three obvious directions for future research. The first would be a direct extension of our work, while the latter two would be tangentially related.

First, as mentioned in the introduction, we could study transport of the charge associated with the current $I_2^{\mu}$ discussed in section \ref{stressenergy}. In particular, the authors of ref. \cite{Son:2009tf} (following the results of refs. \cite{Erdmenger:2008rm,Banerjee:2008th}) showed that associated with any current with an $\EB$ anomaly is a special transport coefficient whose form is fixed by the anomaly coefficient and the equation of state. Our $I_2^{\mu}$ is appears to be such an anomalous current, hence the kinetic coefficient associated with transport of $I_2^{\mu}$ charge should take the form determined in ref. \cite{Son:2009tf}.

Second, we could introduce a thermal gradient into the holographic setup and compute the thermal conductivity and the thermo-electric transport coefficients (called $\alpha_{ij}$ in the introduction) associated with the flavor fields. A further extension would be to work with two coincident D7-branes, and hence two flavors in the SYM theory, and to compute the thermal conductivity and thermo-electric transport coefficients associated with isospin charge. As demonstrated in refs. \cite{Ammon:2008fc,Basu:2008bh,Ammon:2009fe}, a sufficiently large isospin chemical potential triggers a phase transition to a superconducting (more accurately, superfluid) phase, so a holographic study of thermo-electric transport may be relevant for high-$T_c$ superconductors, which exhibit unusually large thermo-electric response even outside the superconducting phase.

Third, we could compute the full conductivity tensor of $\N=4$ SYM theory itself (without flavor), which remains to be done. To date, only the longitudinal conductivity, which we called $\s_{xx}$, has been computed. To compute $\s_{xy}$ and $\s_{xz}$ for $\N=4$ SYM theory holographically would require new supergravity solutions, however. In particular, a nonzero Hall current requires a nonzero density and a nonzero magnetic field, hence we would first need to find a supergravity solution describing a dyonic black hole. Such a solution exists for (3+1)-dimensional AdS, but not yet for (4+1)-dimensional AdS.

\section*{Acknowledgements}
We would like to thank J. Mas, P. Kerner, J. Shock, and K. Skenderis for useful conversations, and J. Erdmenger and A. Karch for reading and commenting on the manuscript. We also thank E. Thompson for collaboration during the early stages of this project. This work was supported in part by the Cluster of Excellence ``Origin and Structure of the Universe.'' A. O'B. would like to thank the Aspen Center for Physics for hospitality during the completion of this project. M. A. would like to thank the Studienstiftung des deutschen Volkes for financial support.

\smallskip
\bigskip

\section*{Appendix: Derivatives of the On-Shell Action}

In this appendix we write explicit expressions for derivatives of the on-shell action with respect to various fields, as mentioned in section \ref{stressenergy}.

For notational simplicity, we first define a function
\beq
d(u) = g_{uu} |g_{tt}| g_{xx}^3 - g_{xx} A_2 - A_4,
\eeq
where $A_2$ and $A_4$ were defined in eqs. (\ref{actiona2def}) and (\ref{actiona4def}), which we repeat here for completeness:
\beq
A_2  = \gu \gx \E^2 + g_{tt} \gu (\Bx^2 + \Bz^2) + \gx^2 \atsq + g_{tt}g_{xx} \left( \axsq + \aysq + \azsq \right),
\eeq
\bea
A_4 & = & \gx \E^2 \left( \aysq + \azsq \right) + \gx \atsq \left( \Bx^2 + \Bz^2 \right) + \gu \E^2 \Bx^2 + g_{tt} \Bz^2 \azsq + g_{tt} \Bx^2 \axsq \\ & & +  2 g_{tt} \Bx \Bz \ax \az - 2 \gx \E \Bz \at \ay. \nonumber
\eea
Recall from section \ref{d7solution} that in our notation $g_{uu}$ represents the $uu$ component of the induced D7-brane metric: $g_{uu} = \frac{1}{u^2} + \theta'(u)^2$.

The derivatives $\frac{\partial L}{\partial F_{\mu\nu}}$, evaluated on our solution, are
\bea
\frac{\partial L}{\partial E_x} & = & \frac{\N \cos^3\theta}{\sqrt{d(u)}} \left [ g_{xx} \Bz \at \ay - \E \left(g_{uu}  \left(g_{xx}^2 + \Bx^2 \right) + g_{xx} \left( \aysq + \azsq \right) \right) \right ], \nonumber \\
\frac{\partial L}{\partial E_y} & = & \frac{\N \cos^3\theta}{\sqrt{d(u)}} \left [  \left( \E \ax \ay + \Bx \at \az - \Bz \at \ax\right ) g_{xx} \right],\nonumber \\
\frac{\partial L}{\partial E_z} & = & \frac{\N \cos^3\theta}{\sqrt{d(u)}} \left [ g_{xx} \E \ax \az -  g_{xx} \Bx \at \ay - g_{uu} \E \Bx \Bz \right ],\nonumber \\
\frac{\partial L}{\partial B_x} & = & \frac{\N \cos^3\theta}{\sqrt{d(u)}} \left [ \Bx \left( g_{uu} |g_{tt}| g_{xx} + |g_{tt}| \axsq - g_{xx} \atsq - g_{uu} \E^2 \right) + |g_{tt}| \Bz \ax \az \right ],\nonumber \\
\frac{\partial L}{\partial B_y} & = & \frac{\N \cos^3\theta}{\sqrt{d(u)}} \left [ |g_{tt}| \Bx \ax \ay + |g_{tt}| \Bz \ay \az - g_{xx} \E \at \az \right ],\nonumber \\
\frac{\partial L}{\partial B_z} & = & \frac{\N \cos^3\theta}{\sqrt{d(u)}} \left [ \Bz \left( g_{uu} |g_{tt}| g_{xx} + |g_{tt}| \azsq - g_{xx} \atsq \right) + |g_{tt}| \Bx \ax \az + g_{xx} \E \at \ay \right ].\nonumber
\eea
The variations with respect to the $\partial_{\mu} \theta$ (with $\mu = t,x,y,z$), evaluated on our solution, are
\bea
\frac{\delta L}{\delta \partial_t \theta} & = & - \frac{\N \cos^3\theta}{\sqrt{d(u)}} \left [ \Bx \Bz \az + \ax \left( g_{xx}^2 + \Bx^2 \right) \right ] \E \, \theta',\nonumber \\
\frac{\delta L}{\delta \partial_x \theta} & = & - \frac{\N \cos^3\theta}{\sqrt{d(u)}} \left [  |g_{tt}| g_{xx} \Bz \ay - \E \at \left( g_{xx}^2 + \Bx^2 \right)\right ] \theta',\nonumber \\
\frac{\delta L}{\delta \partial_y \theta} & = & - \frac{\N \cos^3\theta}{\sqrt{d(u)}} \left [ \Bx \az \left( |g_{tt}| g_{xx} - \E^2 \right) - |g_{tt}| g_{xx}\Bz \ax \right] \theta',\nonumber \\
\frac{\delta L}{\delta \partial_z \theta} & = & + \frac{\N \cos^3\theta}{\sqrt{d(u)}} \left [  \E \Bz \at + \ay \left( |g_{tt}| g_{xx} - \E^2 \right) \right] \Bx \, \theta'.\nonumber
\eea

\bibliographystyle{JHEP}
\bibliography{hall2}

\end{document}